\documentclass[twocolumn, prb,showpacs]{revtex4-1}

\usepackage{amsmath}
\usepackage{mathrsfs} 
\usepackage{amssymb} 
\usepackage{graphicx}
\usepackage{subfigure}
\usepackage{dcolumn}
\usepackage{bm}
\usepackage{natbib}

\begin{document}

\title{Tuning the electronic structures of silicene and germanene by biaxial strain and electric field }
\date{\today}
\author{Jia-An Yan$^1$}
\email{jiaanyan@gmail.com}
\affiliation{1. Department of Physics, Astronomy, and Geosciences, Towson University, 8000 York Road, Towson, MD 21252, USA\\
2. Department of Materials Science, Fudan University, 220 Han Dan Lu, Yangpu, Shanghai 200433, China}
\author{Shang-Peng Gao$^2$}
\affiliation{1. Department of Physics, Astronomy, and Geosciences, Towson University, 8000 York Road, Towson, MD 21252, USA\\
2. Department of Materials Science, Fudan University, 220 Han Dan Lu, Yangpu, Shanghai 200433, China}
\author{Ryan Stein$^1$}
\affiliation{1. Department of Physics, Astronomy, and Geosciences, Towson University, 8000 York Road, Towson, MD 21252, USA\\
2. Department of Materials Science, Fudan University, 220 Han Dan Lu, Yangpu, Shanghai 200433, China}
\author{Gregory Coard$^1$}
\affiliation{1. Department of Physics, Astronomy, and Geosciences, Towson University, 8000 York Road, Towson, MD 21252, USA\\
2. Department of Materials Science, Fudan University, 220 Han Dan Lu, Yangpu, Shanghai 200433, China}

\begin{abstract}
We present a first-principles study of effects of small biaxial strain ($|\varepsilon|\le 5\%$) and perpendicular electric field (E-field) on the electronic and phonon properties of low-buckled silicene and germanene. With an increase of the biaxial strain, the conduction bands at the high symmetric $\Gamma$ and $M$ points of the first Brillouin zone shift significantly towards the Fermi level in both silicene and germanene. In contrast, the E-field changes the band dispersions near the $\Gamma$ and open a small band gap at the K point in silicene. We found that the field-induced gap opening in silicene could be enhanced by a compressive strain while mitigated by a tensile strain. This result highlights the tunability of the electronic structures of silicene by combining the mechanical strain and the electric field.
\end{abstract}

\maketitle

\section{Introduction}

The success of graphene has stimulated great interest in novel two-dimensional (2D) atomic crystals for interesting physics and diverse applications \cite{Novoselov2005}. Silicene and germanene, the silicon and germanium analogs of graphene, are currently attracting growing attention \cite{Takeda1994,Cahangirov2009,Verri2007, Ezawa2012, Kara2012}. Similar to graphene, both of these crystals are shown to exhibit linear band dispersions near the Fermi level at the K point of the first Brillouin zone (BZ) in the absence of spin-orbit coupling (SOC)\cite{Cahangirov2009}. Due to their low-buckled structure, the sublattice inversion symmetry can be broken by applying a perpendicular electric field (E-field), leading to a sizable band gap of up to tens of meV \cite{Ni2012,Drummond2012,Ezawa2012njp}. Such a field-tunable band gap is useful for possible electronic device applications. The recent experimental fabrication of silicene on different substrates \cite{Vogt2012,Fleurence2012,Chen2012,Feng2012,Jamgotchian2012,Lin2012,Paola2014,Yamada2014} has facilitated to further probe the material and exploit its properties.

Since silicene and germanene are most likely to be fabricated on substrates \cite{Fleurence2012,Vogt2012,Chen2012,Feng2012,Jamgotchian2012,Lin2012}, external strain induced by lattice constant mismatch between silicene/germanene and substrate may be present. In a recent theoretical proposal \cite{Cai2013}, silicene may be grown on graphene, with a compressive strain up to $\varepsilon = -3 \%$ on the silicene lattice. Therefore, understanding of the strain effect on the electronic and phonon properties will be essential for silicene-based applications \cite{Huang2013,Balandin2012}. On the other hand, the in-plane strain provides an independent degree of freedom to tune the electronic properties in these 2D crystals \cite{Qin2012,Hu2013,Kaloni2013}. In fact, strain has already become a powerful means of tailoring the electronic structure and to affect the carrier mobility in silicon-based materials \cite{Hoyt2002,Thompson2004} and in two-dimensional graphene \cite{Pereira2009, Levy2010, SBL2013}. It has been reported that the in-plane biaxial and uniaxial strains could dramatically change the Fermi velocities in silicene \cite{Qin2014}. Small biaxial strain of 5\% also opens up channels possibly useful for an enhanced electron-phonon coupling in silicene \cite{Wan2013}. Furthermore, strain may introduce delicate change on the phonon dispersions, dictating interesting Raman spectra as observed in graphene \cite{Mohiuddin2009,Metzger2010,Ding2010,Zabel2012,Ferralis2010}.

It is expected that by combining an in-plane strain with a perpendicular E-field, a wider range of tunability on the electronic structure of silicene and germanene may be possible. How will these two factors affect each other? To address this question, we have carried out detailed first-principles calculations to study electronic properties of silicene and/or germanene with small biaxial strain (with ($|\varepsilon|\le$5\%)) under an additional E-field up to 5 V/nm. A possible E-field strength of up to 3 V/nm in bilayer graphene has been realized using dual-gated structure in experiment\cite{Zhang2009}. We limit our study to this small strain range so that silicene and germanene will not alter the semimetallic nature. Besides, such a small strain will not change the crystal structure dramatically, and achievable in experiments \cite{Huang2012, Li2012}. Despite of recent work on silicene under strain \cite{Qin2012,Hu2013,Kaloni2013,Qin2014}, a first-principles study including \emph{both} strain and E-field in silicene has not been reported.

The paper is organized as follows. Calculational details are given in Section II. In Section III.A, we will first consider the strain effect on the crystal structure of silicene and germanene. The strain effects on the band structures and phonon dispersions of silicene and germanene are presented in III.B and III.C. We then focus our discussions on silicene with both strain and E-field in Section III.D. A summary is given in Section IV.

\begin{figure}[tbp]
\centering
   \includegraphics[width=7.0cm,clip]{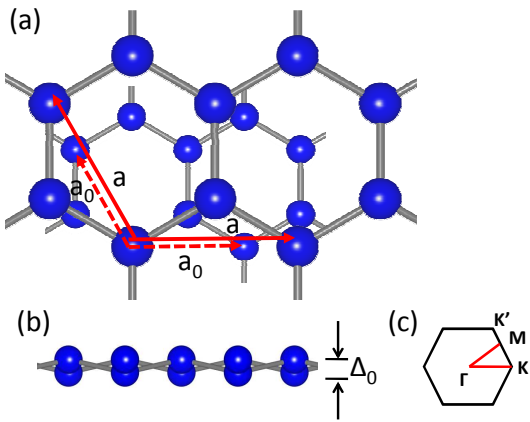}
 \caption{(Color online) (a) Top and (b) side views of the low-buckled silicene crystal lattice. The lattice constants of unstrained and strained silicene are denoted as $a_0$ and $a$, respectively. The buckling constant $\Delta_0$ is shown in (b). (c) The high symmetric points $\Gamma$, K, and M in the first BZ of silicene. }\label{fig1}
\end{figure}

\section{Methods}

Our first-principles calculations were performed using density functional theory (DFT) and density-functional perturbation theory (DFPT) \cite{Baroni2001} with the local density approximation (LDA) as implemented in the Quantum ESPRESSO code \cite{pwscf}. Norm-conserving pseudopotentials \cite{Troullier1991} for Si and Ge were adopted to describe the core-valence interactions. The wave functions of the valence electrons were expanded in plane waves with a kinetic-energy cutoff of 36 and 40 Ry for silicene and germanene, respectively. A Monkhorst-Pack uniform $k$-grid of 36$\times$36$\times$1 is used to do the self-consistent calculations. For the phonon dispersions, we used a 6$\times$6$\times$1 $q$-grid. A vacuum region of 20 \AA~ is introduced to eliminate any artificial interaction between neighboring supercells along the perpendicular direction. The relaxed lattice constants are $a_0$ = 3.83 \AA~ for silicene and 3.95 \AA~ for germanene. The relaxed buckled separations are $\Delta_0$ = 0.44 \AA~for pristine silicene and 0.64 \AA~for pristine germanene. These results are in good agreement with previous DFT calculations \cite{Cahangirov2009,Bechstedt2012}. After we obtained the structure of unstrained silicene and germanene, various biaxial strains were applied to the siliecene and germanene lattices, respectively, with the biaxial strain $\varepsilon$ defined as $\varepsilon$ = $(a-a_0)/a_0$. Here $a$ and $a_0$ are the strained and unstrained lattice constants, respectively, as schematically indicated in Fig.~\ref{fig1}. According to this definition, a negative $\varepsilon$ means the compressive strain, while a positive value indicates the tensile stress. For each strain, the atomic positions were fully relaxed until the force is smaller than 0.02 eV/\AA.

\section{Results}

\subsection{ Effects of strain on the buckling constants}

\begin{figure}[tbp]
\centering
   \includegraphics[width=7.5cm,clip]{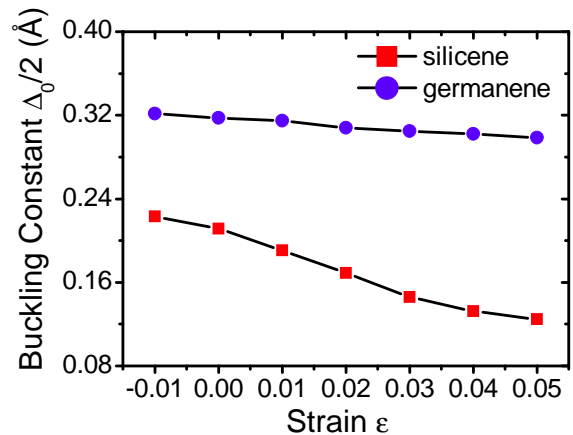}
 \caption{(Color online) Half of the buckling constant $\Delta_0$/2 as a function of applied strain $\varepsilon$ in silicene and germanene. }\label{fig2}
\end{figure}

In silicene and germanene, the low-buckled crystal structure is the main feature distinct from the planar graphene. This buckling is characterized by the vertical distance between the two silicon atoms in the unit cell, i.e., the buckling constant $\Delta_0$, as indicated in Fig.~\ref{fig1}(b). It is expected that an in-plane tensile strain decreases $\Delta_0$. Figure~\ref{fig2} shows $\Delta_0$/2 as a function of $\varepsilon$ in both silicene and germanene for $\varepsilon$ =-0.01$\sim$ 0.05. When a positive biaxial strain is applied, the lattice constant increases. As $\varepsilon$ increases from 0 to 0.05, $\Delta_0$/2 decreases 41\% from 0.22 to 0.12 \AA~in silicene. In contrast, the buckling constant decreases only 6\% in germanene for the same strain range, much smaller than in silicene. This is mainly due to the distinct atomic radii of Si and Ge. The single-bond radii are 1.09 \AA~for silicon and 1.22 \AA~for Ge, respectively. The larger atomic radius of Ge makes the Ge-Ge bond insensitive to a relatively small strain. These results are overall consistent with previous DFT calculations\cite{Kaloni2013}.

\subsection{Effects of strain on the band structures}

\begin{figure*}[th]
\centering
   \includegraphics[width=12 cm, angle = -90, clip]{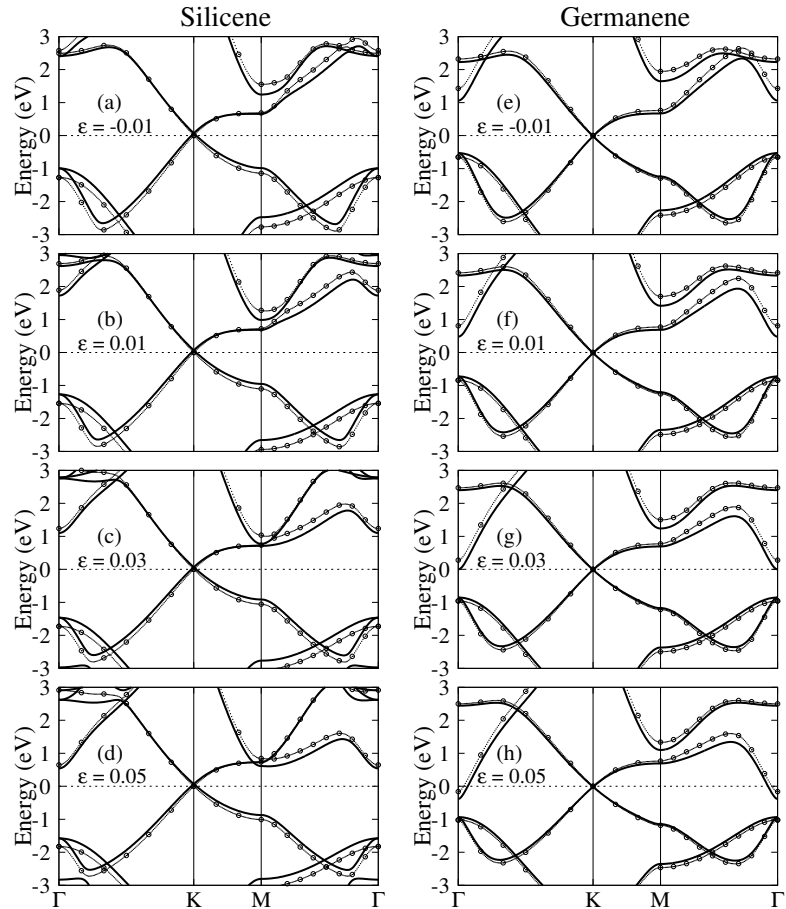}
 \caption{Band structures of silicene (left column) and germanene (right column) as a function of strain. The Fermi level has been shifted to zero. Solid line: DFT-LDA band dispersions; Dotted line with symbols: interpolated GW band structures.
 }\label{fig3}
\end{figure*}

\begin{figure}[tbp]
\centering
   \includegraphics[width=7.0cm,clip]{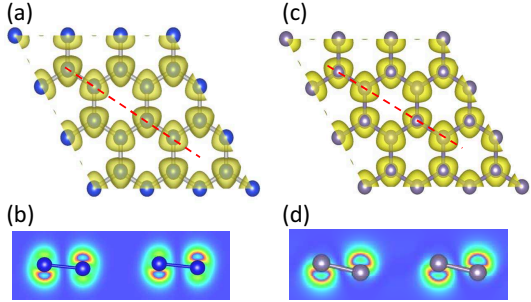}
 \caption{(Color online) Charge density distribution for the CBM at $\Gamma$ with strain $\varepsilon$ = 0.05. (a) Isosurface and (b) cross-sectional view of the charge distribution in silicene; (c) Isosurface and (d) cross-sectional view of the charge density distribution in germanene. The cross-section planes are indicated by dashed line in (a) and (c), respectively.}\label{fig:chg}
\end{figure}

The biaxial strain has dramatic effects on the electronic properties of 2D crystals. Figure~\ref{fig3} shows the band structures of silicene and germanene as a function of strain $\varepsilon$ from -0.01 to 0.05. In this range, a few interesting features can be summarized for silicene. First, the linear band dispersions near the K point keep almost intact for both silicene and germanene. Second, the biaxial strain shows more evident effects on the bands at the $\Gamma$ and M points than at the K point. In particular, the conduction band minimum (CBM) moves towards the Fermi level as $\varepsilon$ increases. The CBM decreases from 2.4 eV for $\varepsilon$ = -0.01 (Fig.~\ref{fig3}(a)) to 0.5 eV for $\varepsilon$ = 0.05 (Fig.~\ref{fig3}(d)) in silicene. This band also exhibits anisotropic dispersions along the $\Gamma$-M and $\Gamma$-K directions. In contrast, the valence band maximum (VBM) at $\Gamma$ moves downward into deeper energy by the tensile strain. In Fig.~\ref{fig:chg} we show the isosurface of the charge density for the CBM at $\Gamma$ in both silicene and germanene. Interestingly, this band exhibits somewhat $\pi$-like bonding features.

The biaxial strain has significant effects on the parabolic conduction bands near the M point. The parabolic bands moves from 1.2 eV for $\varepsilon$ = -0.01 (Fig.~\ref{fig3}(a)) to 0.7 eV from $\varepsilon$ = 0.05 (Fig.~\ref{fig3}(d)) in silicene. The parabolic band begins to cross with the $\pi^*$ band near the M point, as shown in Fig.~\ref{fig3}(d).

Comparing to silicene, the strain shows similar but more evident effects on the band dispersions in germanene. The CBM at $\Gamma$ is very sensitive to the strain. For $\varepsilon$ = 0.03, the CBM at $\Gamma$ already touches the Fermi level, leading to semimetal-to-metal transition, as shown in Figs.~\ref{fig3} (g) and (h). Similar result has also been reported for silicene with larger tensile strain \cite{Kaloni2013}.

The strain-induced shift on the parabolic conduction bands (band No. 6) at the M point changes the band separations and consequently opens new scattering channels for the electron-phonon coupling (EPC). It has been reported that there will be a significant enhancement of the EPC in electron-doped strained silicene \cite{Wan2013}. To see whether the sing-particle band shift induced by the strain is still valid after including the many-body effects, we have performed GW calculations using the ABINIT code \cite{Gonze1988}. Following the standard approach of one-shot G$_0$W$_0$ method, Kohn-Sham eigenvalues and eigenfunctions were firstly obtained by DFT-LDA calculation and then used as a starting point to do the GW correction \cite{Gonze1988,Hybertsen1986,Godby1988}. Due to the non-local nature of the self-energy operator, the K-point used for the GW calculation has to belong to the K-point grid that has been chosen for the Kohn-Sham self-consistent calculation \cite{Gonze1988}. A 18$\times$18$\times$1 Monkhorst-Pack grid is used. The GW band structures were interpolated based on the GW energies explicitly calculated at the grid and the DFT-LDA band dispersions. The screening in the GW calculation is treated with the plasmon-pole model \cite{Hybertsen1986,Gonze1988}. The polarization function is calculated within the random phase approximation. Coulomb interaction is truncated using a cut-off radius of 5 \AA~along the perpendicular direction. The number of bands used to calculate the screening and the self-energy in the GW method is chosen to be 138. The cut-off energy of the plane waves is set to 327 eV (24 Ry) to represent the independent particle susceptibility and 381 eV (28 Ry) to represent the dielectric matrix and to generate the exchange part of the self-energy operator. The results have been shown in Fig.~\ref{fig3}.

From Fig.~\ref{fig3}, one can see that overall the GW band structures follow the DFT-LDA trend, i.e., the parabolic bands at the M point shift towards the Fermi level as $\varepsilon$ increases. The GW quasiparticle bands also exhibit larger Fermi velocity near the K point, in agreement with previous report \cite{Huang2013}. From Fig.~\ref{fig3}, the band separations will become to match the phonon energies for even smaller tensile strain. Different from the single-particle picture \cite{Wan2013}, one may expect that the EPC enhancement will occur at an even smaller tensile strain.

\subsection{Effects of strain on the phonon dispersions}

\begin{figure}[th]
\centering
   \includegraphics[width=9.0cm,clip]{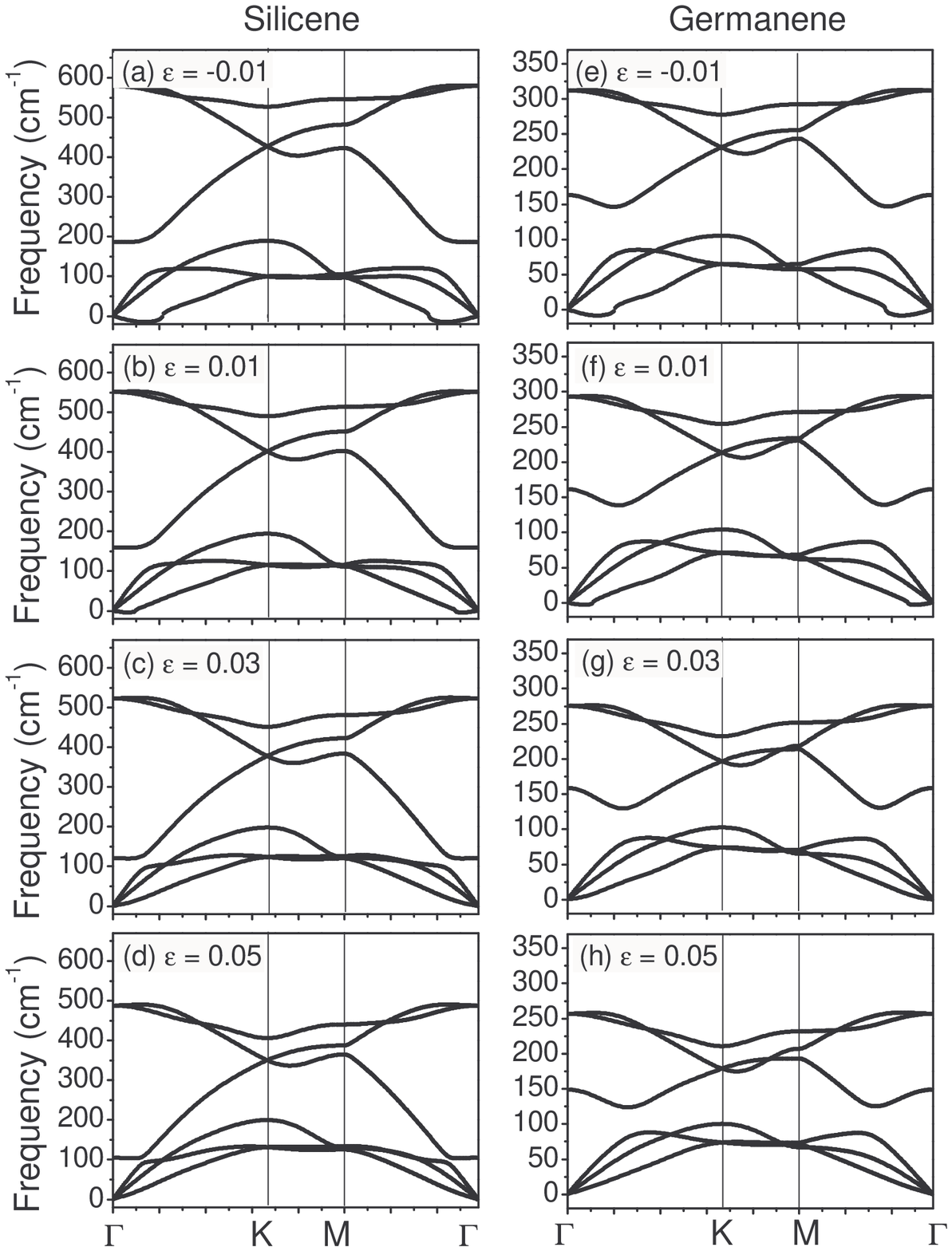}
 \caption{Phonon dispersions of silicene and germanene as a function of strain $\varepsilon$. Left column: silicene; Right column: germanene.}\label{fig:phband}
\end{figure}

The effects of biaxial strain on the phonon dispersions in silicene and germanene are shown in Fig.~\ref{fig:phband}. Note that in Figs.~5(a)-(b) and 5(e)-(f), small negative phonon frequencies (within -20 cm$^{-1}$) appear on the acoustic branch along the $\Gamma$-K direction. These small negative modes, however, do not mean the mechanical instability of the system. Instead, it mainly comes from the interpolation process as implemented in the Quantum Espresso code \cite{pwscf}. Hence, both silicene and germanene will be stable with the biaxial strain $\varepsilon$ from -0.01 to 0.05. Generally, isotropic compression results in phonon stiffening of the vibrational mode, while isotropic tension results in a decrease on the vibrational frequency (phonon softening). From Fig.~\ref{fig:phband}, the tensile strain lowers the vibrational frequency, especially the in-plane and out-of-plane optical modes in silicene and germanene.

The rate of change in frequency as a function of strain for a given phonon mode in a crystal is determined by its Gr\"{u}neisen parameter $\gamma_g$. It is a crucial parameter to quantify the rate of the phonon mode softening (stiffening) under tensile (compressive) strain and determines the thermo-mechanical properties. In metrology applications, accurate determination of the Gr\"{u}neisen parameters are key for quantifying the amount of strain in the system. In presence of biaxial strain, the Gr\"{u}neisen parameter $\gamma_g$ for a particular band $m$ associated with an in-plane phonon mode reads\cite{Mohiuddin2009,Ferralis2010,Cheng2011}:
\begin{equation}
\gamma_g = -\frac{1} {2\omega_m^0} \frac{\partial \omega_m}{\partial \varepsilon},
\end{equation}
where $\varepsilon$ is the biaxial strain applied to the system, and $\omega_m^0$ and $\omega_m$ correspond to the phonon frequencies at zero strain and in presence of an applied strain, respectively.

We have performed a linear fitting of the phonon frequency shifts as a function of $\varepsilon$. The Gr\"{u}neisen parameters are then calculated using Eq.~(1) based on the obtained linear slopes. Table~I summarizes the results. For comparison, we also listed the calculational results for graphene. The calculated $\gamma_g$ for the $\Gamma-E_{2g}$ and K-$A_1'$ modes in graphene are 1.8 and 2.8, in good agreement with the values of 1.8 and 2.7 obtained by Mohiuddin \emph{et al.} \cite{Mohiuddin2009}. Surprisingly, silicene has the smallest $\gamma_g$ of 1.3 for the $\Gamma-E_g$ mode, which is 14\% smaller than in germanene and 25\% smaller than in graphene. Our LDA value is smaller than the GGA result of 1.6 obtained by Kaloni \emph{et al.} \cite{Kaloni2013}. As can be seen from Table I, the $\gamma_g$ for the K-$A$ mode are nearly 2.8, 2.0 and 2.1 for graphene, silicene, and germanene, respectively. We conclude that the 2D band in the Raman spectra, i.e., the overtone of the D mode (corresponding to the K-$A$ mode), may be more sensitive to the biaxial strain than the G band.

\begin{figure}[th]
\centering
   \includegraphics[width=7.0cm,clip]{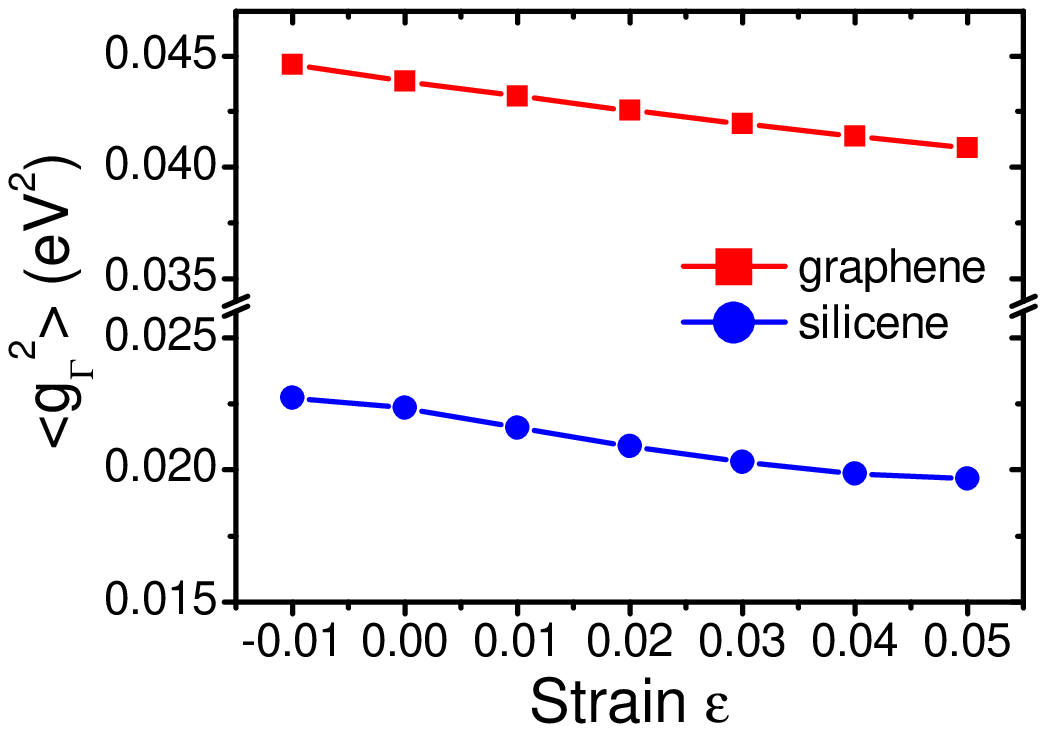}
 \caption{EPC square as a function of strain in silicene and graphene.}\label{fig:epc}
\end{figure}

Finally, we consider the effects of strain on the EPC for the $\Gamma-E_g$ mode in silicene. In our previous work \cite{me2013}, the EPC in germanene is found to be an order of magnitude smaller than in silicene. Therefore, we only focus on silicene and compare the results with graphene. Since at $k$ = K, the electronic states are doubly degenerate, we calculate the average EPC square over the Fermi surface defined as $\langle g_q^2\rangle_F = \sum_{i,j}^\pi |g_{(K+q)i,Kj}|^2/4$ with $q$ = $\Gamma$, where the sum is performed over the doubly degenerate $\pi$ bands at $E_F$. This quantity is found to be a good indication of the EPC strength of the $E_g$ mode for the electronic states near the BZ corner. In silicene, the calculated value is $\langle g_\Gamma^2\rangle_F = \sum_{i,j}^\pi |g_{Ki,Kj}|^2/4$ = 0.0223 eV$^2$. As depicted in Fig.~\ref{fig:epc}, we found that the EPC square decreases as $\varepsilon$ increases. At $\varepsilon$ = 0.05, the EPC square in silicene is 0.0197 eV$^2$, about 12\% smaller than in unstrained silicene. In graphene, the EPC square is 0.0409 eV$^2$, nearly 7\% smaller than the unstrained one. This result shows that the EPC for the $\Gamma-E_g$ mode in silicene is almost two times more sensitive to the biaxial strain than in graphene.


\begin{table}[tbp]
 \caption{Calculated Gr\"{u}neisen parameters for the highest optical modes at $\Gamma$ and at $K$ in graphene, silicene and germanene, respectively.} \label{tab}
\begin{ruledtabular}
\begin{tabular}{lcccccc}
            &   \multicolumn{2}{c}{Graphene}  & \multicolumn{2}{c}{Silicene} & \multicolumn{2}{c}{Germanene}\\
            \cline{2-3} \cline{4-5} \cline{6-7}
            & $\Gamma-E_{2g}$ & $K-A_1'$ & $\Gamma-E_g$ & $K-A$ & $\Gamma-E_g$ & $K-A$ \\
            \hline
$\gamma_g$  & 1.75, 1.8\footnote{Ref. \citenum{Mohiuddin2009}.} & 2.75, 2.7$^{a}$ & 1.31, 1.64\footnote{Ref. \citenum{Kaloni2013}.} & 1.95 & 1.52 & 2.11\\

\end{tabular}
\end{ruledtabular}
\end{table}

\subsection{Effects of both strain and E-field on the band structures}

The perpendicular E-field has been shown to open a small energy gap in silicene \cite{Ni2012,Drummond2012}. This feature makes silicene a very interesting platform for electrically tunable device applications. However, the required E-field is too strong (up to 5 V/nm), and the opened band gap is relatively small ($\sim$ 20 meV), which limits the practical application of silicene. Will the biaxial strain possibly enhance the gap opening? Since the tunable strain range is relatively small in germanene, in this part we will mainly focus on the effects of \emph{both} the E-field and the biaxial strain on the electronic structure of silicene.

\begin{figure}[tbp]
\centering
   \includegraphics[width=9.0cm,clip]{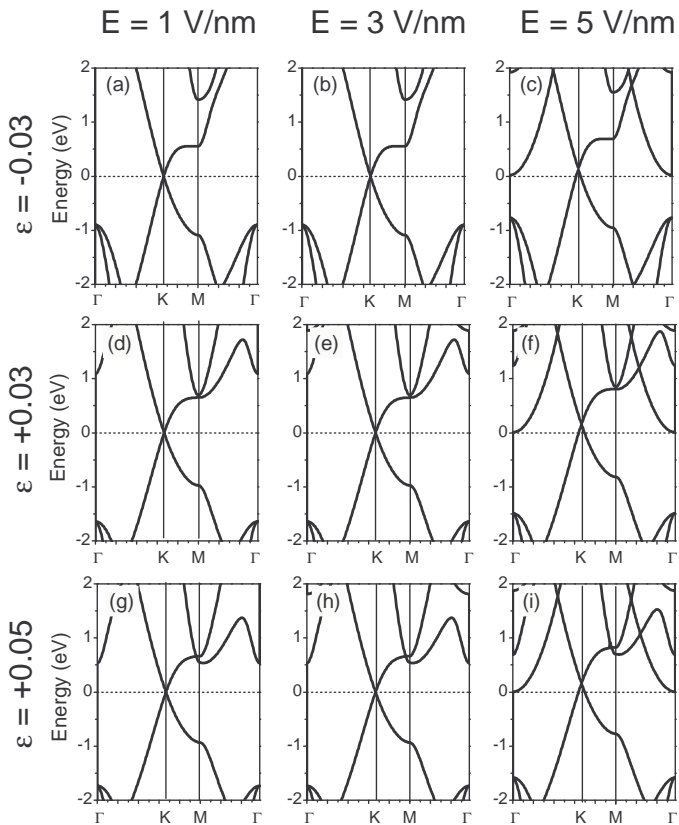}
 \caption{Band structures of silicene under various E-fields for typical strains. $\varepsilon$ = -0.03 with $E$ = 1 V/nm (a), 3 V/nm (b), and 5 V/nm (c). $\varepsilon$ = +0.03 with $E$ = 1 V/nm (d), 3 V/nm (e) and 5 V/nm (f). $\varepsilon$ = +0.05 with $E$ = 1 V/nm (g), 3 V/nm (h) and 5 V/nm (i). }\label{fig:band-field}
\end{figure}

Figure~\ref{fig:band-field} presented the band dispersions of silicene under various E-field ($E$ = 1, 3, and 5 V/nm) for three typical strains ($\varepsilon$ = -0.03, +0.03 and +0.05). When silicene is under a given strain $\varepsilon$, the additional perpendicular E-field shows two main effects: (1) The E-field moves the CBM at $\Gamma$ down towards the Fermi level; and (2) the E-field opens a small band gap at K. The CBM at $\Gamma$ will approach the Fermi level and eventually a semimetal-to-metal transition takes place at a high enough E-field (see Fig.~\ref{fig:band-field}(a)-(c)). Meanwhile, the energy band gap opened by the E-field also increases as the field strength becomes stronger, as will be shown below.

\begin{figure}[tbp]
\centering
   \includegraphics[width=6.5cm,clip]{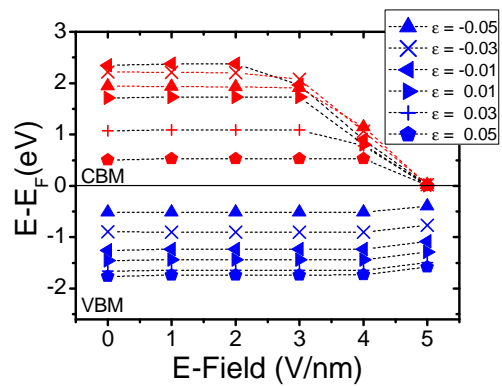}
 \caption{(Color online) Evolution of the CBM and VBM at the $\Gamma$ point as a function of the E-field for various strain $\varepsilon$ in silicene. Red symbols denote CBM, while blue symbols represent VBM. }\label{fig:cbm}
\end{figure}

The effects of strain and E-field can be seen more clearly from Fig.~\ref{fig:cbm}, where the evolutions of the CBM and VBM at the $\Gamma$ point are plotted as a function of the E-field for various strain $\varepsilon$. Interestingly, the VBM doesn't show any evident dependence on the E-field until 5 V/nm. In contrast, the biaxial strain introduces rigid shifts on the VBM. In particular, a compressive strain shifts the VBM towards the Fermi level.

On the other hand, the CBM at $\Gamma$ exhibits a clear dependence on both strain and the E-field, as shown in Fig.~\ref{fig:band-field}. For a given E-field, the CBM shifts to higher energy from $\varepsilon$ = -0.05 to -0.01 and then moves back from $\varepsilon$ = -0.01 to 0.05. For silicene with a specific strain $\varepsilon$, the increase of the E-field will push the CBM downwards and touch the Fermi level at $E$ = 5 V/nm. Moreover, the compressive strain seems to further enhance such a shift at $\Gamma$.

\begin{figure}[tbp]
\centering
   \includegraphics[width=9.0cm,clip]{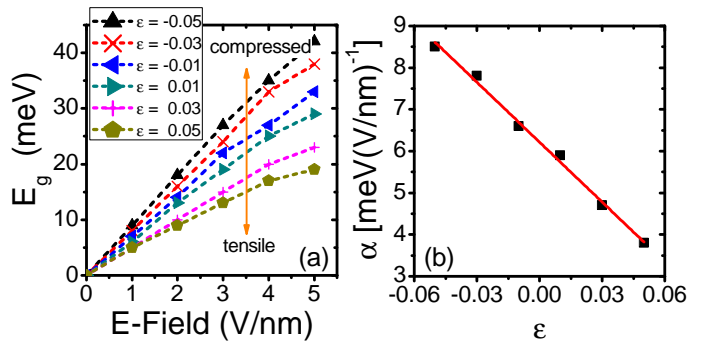}
 \caption{(Color online) (a) Energy band gap $E_g$ at the K point as a function of the E-field for various strain $\varepsilon$ in silicene. (b) The calculated enhancement factor $\alpha$ as a function of strain $\varepsilon$. }\label{fig:gap}
\end{figure}

In addition to the $\Gamma$ point, the co-application of both the E-field and the strain in silicene has dramatic effects on the band gap opening at K. Figure~\ref{fig:gap} summarizes the results.

The energy gap $E_g$ at K changes as a function of both the E-field ($E$ = 0 $\sim$ 5 V/nm) and the biaxial strain $\varepsilon$ (from -0.05 to 0.05), as shown in Fig.~\ref{fig:gap}(a). For a given strain, the $E_g$ at K increases almost linearly as the E-field increases. For example, for $\varepsilon$ = 0.01, the $E_g$ increases from 0 to 30 meV when $E$ changes from 0 to 5 V/nm. In contrast, the increase of the $E_g$ becomes saturated quickly as $\varepsilon$ changes to tensile strain. On the other hand, the compressive strain significantly enhances the band gap opening. For $\varepsilon$ = -0.05, the band gap changes from 0 to 42 meV at $E$ = 5 V/nm, more than 35\% larger than that of $\varepsilon$ = 0.01 \cite{note}.

The above behavior may be understood from the strain effect on the buckling constant $\Delta_0$, as shown in Fig.~\ref{fig2}. The compressive strain enlarges the vertical distance of two silicon atoms, and therefore, increases the on-site energy difference between two sublattices. Using two-band model at the K point, one can write out the Hamiltonian:
\begin{equation}
H =
\left(
  \begin{array}{cc}
    +eE\Delta_0/2 & 0\\
    0             &-eE\Delta_0/2  \\
  \end{array}
\right),
\end{equation}
with $E$ the field strength and $\Delta_0$ the buckling constant. The energy band gap is given by $E_g = eE\Delta_0$. From Fig.~\ref{fig2}, the compressive (tensile) strain increases (decreases) the $\Delta_0$, and consequently, enhances (mitigates) the gap opening. As field strength increases, the screening becomes stronger, especially for silicene under tensile strain with smaller buckling constant. This will decrease the displacement field strength and as a result, the band gap becomes saturated quickly, as shown in Fig.~\ref{fig:gap}. This is similar to the results observed in bilayer graphene \cite{Zhang2009}.

One can quantify the effect of strain on the gap opening. Since $E_g$ changes linearly with $E$, we define the enhancement factor $\alpha$ by fitting to the expression $E_g = \alpha(\varepsilon) E$ for each $\varepsilon$. Here $\alpha (\varepsilon)$ is a function of $\varepsilon$. $E$ is the field strength in V/nm, and $E_g$ the energy band gap in meV. In Fig.~\ref{fig:gap}(b), the calculated $\alpha$ is plotted as a function of $\varepsilon$. Clearly, $\alpha$ decreases almost linearly with respect to the biaxial strain $\varepsilon$. This relationship can be described as $\alpha$ = 6.2-47.8$\varepsilon$. For the compressive strain of $\varepsilon$ = -0.05, we have $\frac{\alpha(-0.05)}{\alpha(0)}$ = 1.4, while for the tensile strain of $\varepsilon$ = 0.05, the ratio is $\frac{\alpha(0.05)}{\alpha(0)}$ = 0.6. Clearly, for a given E-field, the compressive strain will enhance the gap opening by increasing $\alpha$, while a tensile strain mitigates this trend. This result highlights the impact of the strain on the field-induced gap opening in silicene.

\section{Summary}

In summary, we have performed detailed first-principles calculations to study the effects of the biaxial strain and the E-field on the electronic and phonon properties of low-buckled silicene and germanene. The small biaxial strain is found to dramatically change the conduction bands at $\Gamma$ and M, while the E-field mainly affects the bands at the $\Gamma$ and K points. The Gr\"{u}neisen parameters for the $\Gamma-E_g$  and the K-$A$ mode in silicene and germanene are calculated and compared with graphene. We also show that the EPC matrix square for the $\Gamma-E_g$ mode in silicene is more sensitive to the biaxial strain than in graphene. Finally, the field-induced band gap is found to be sensitive to the strain. In particular, the compressive strain is able to significantly enhance the field-induced gap opening in silicene.

\begin{acknowledgments}
J.A.Y. and R.S. were supported by the SET funding at the Towson University. J.A.Y. also acknowledges the Faculty Development and Research Committee grant (OSPR No. 140269) and the FCSM Fisher General Endowment at the Towson University. S.P.G. is supported by the State Key Development Program of Basic Research of China (Grant No. 2011CB606406). The computational resources utilized for the GW calculation in this research are provided by Shanghai Supercomputer Center.
\end{acknowledgments}


\begin{thebibliography}{99}

\bibitem{Novoselov2005} K. S. Novoselov, D. Jiang, F. Schedin, T. J. Booth, V. V. Khotkevich, S. V. Morozov, and A. K. Geim, Proc. Natl Acad. Sci. USA \textbf{102}, 10451 (2005).

\bibitem{Takeda1994} K. Takeda and K. Shiraishi, Phys. Rev. B \textbf{50}, 14916-14922 (1994).

\bibitem{Cahangirov2009} S. Cahangirov, M. Topsakal, E. Akt\"{u}rk, H. Sahin, and S. Ciraci, Phys. Rev. Lett. \textbf{102}, 236804 (2009).

\bibitem{Verri2007} G. G. Guzm\'{a}n-Verri, L. C. Lew Yan Voon, Phys. Rev. B \textbf{76}, 075131 (2007).

\bibitem{Ezawa2012} M. Ezawa, Phys. Rev. Lett. \textbf{109}, 055502 (2012).

\bibitem{Kara2012} A. Kara, H. Enriquezc, A. P. Seitsonen, L. C. Lew Yan Voon, S. Vizzini, B. Aufray, H. Oughaddou, Surf. Sci. Rep. \textbf{67}, 1 (2012).


\bibitem{Ni2012} Z. Ni, Q. Liu, K. Tang, J. Zheng, J. Zhou, R. Qin, Z. Gao,
D. Yu, and J. Lu, Nano Lett. \textbf{12}, 113-118 (2012).

\bibitem{Drummond2012} N. D. Drummond, V. Z\'{o}lyomi, and V. I. Fal'ko, Phys. Rev. B \textbf{85}, 075423 (2012).


\bibitem{Ezawa2012njp} M. Ezawa, New J. Phys. \textbf{14}, 033003 (2012).


\bibitem{Vogt2012} P. Vogt, P. De Padova, C. Quaresima, J. Avila, E. Frantzeskakis, M. C. Asensio, A. Resta, B. Ealet,
and G. Le Lay, Phys. Rev. Lett. \textbf{108}, 155501 (2012).


\bibitem{Fleurence2012} A. Fleurence, R. Friedlein, T. Ozaki, H. Kawai, Y. Wang, and Y. Yamada-Takamura, Phys. Rev. Lett. \textbf{108}, 245501 (2012).

\bibitem{Chen2012} L. Chen, C.-C. Liu, B. Feng, X. He, P. Cheng, Z. Ding, S. Meng, Y. Yao, K. Wu, Phys. Rev. Lett. \textbf{109}, 056804 (2012).

\bibitem{Feng2012} B. Feng, Z. Ding, S. Meng, Y. Yao, X. He, P. Cheng, L. Chen, and K. Wu, Nano Lett., \textbf{12}, 3507-3511 (2012).

\bibitem{Jamgotchian2012} H. Jamgotchian, Y. Colignon, N. Hamzaoui, B. Ealet, J. Y. Hoarau, B. Aufray, and J. P. Bib\'{e}rian, J. Phys. Condens. Matter \textbf{24}, 172001 (2012).


\bibitem{Lin2012} C.-L. Lin, R. Arafune, K. Kawahara, N. Tsukahara, E. Minamitani, Y. Kim, N. Takagi, and M. Kawai, App. Phys. Exp. \textbf{5}, 045802 (2012).

\bibitem{Paola2014} Paola De Padova, C. Ottaviani, C. Quaresima, B. Olivieri, P. Imperatori, E. Salomon, T. Angot, L. Quagliano, C. Romano, A. Vona, M. Muniz-Miranda, A. Generosi, B. Paci and Guy Le Lay, 2D Mater. \textbf{1}, 021003 (2014).


\bibitem{Yamada2014} Y. Yamada-Takamura, and R. Friedlein, Science and Technology of Advanced Materials \textbf{15}, 064404 (2014).


\bibitem{Cai2013} Yongmao Cai, Chih-Piao Chuu, C. M. Wei, and M. Y. Chou, Phys. Rev. B \textbf{88}, 245408 (2013).


\bibitem{Huang2013} S. Huang, W. Kang, and L. Yang, Appl. Phys. Lett. \textbf{102}, 133106 (2013).

\bibitem{Balandin2012} A. A. Balandin and D. L. Nika, Materials Today \textbf{15}, 266 (2012).
%

\bibitem{Qin2012} R. Qin, C.-H. Wang, W. Zhu, and Y. Zhang, AIP Advances \textbf{2}, 022159 (2012).

\bibitem{Hu2013} M. Hu, X. Zhang, and D. Poulikakos, Phys. Rev. B \textbf{87}, 195417 (2013)

\bibitem{Kaloni2013} T. P. Kaloni, Y. C. Cheng, and U. Schwingenschl\"{o}gl, J. Appl. Phys. \textbf{113}, 104305 (2013). 


\bibitem{Hoyt2002} J. L. Hoyt, H.M. Nayfeh, S. Eguchi, I. berg, G. Xia, T. Drake, E.A. Fitzgerald, D.A. Antoniadis, IEDM Techn. Digest \textbf{2002}, 23-26 (2002).

\bibitem{Thompson2004} S. E. Thompson, M. Armstrong C. Auth, S. Cea, R. Chau, G. Glass, T. Hoffman, J. Klaus, Z. Ma, B. Mcintyre, A. Murty, B. Obradovic, L. Shifren, S. Sivakumar, S. Tyagi, T. Ghani, K. Mistry, M. Bohr, Y. El-Mansy, , IEEE ED Letters \textbf{25}, 191-193 (2004).

\bibitem{Pereira2009} V. M. Pereira and A. H. Castro Neto, Phys. Rev. Lett. \textbf{103}, 046801 (2009).


\bibitem{Levy2010} N. Levy, S. A. Burke, K. L. Meaker, M. Panlasigui, A. Zettl1, F. Guinea, A. H. Castro Neto, M. F. Crommie, Science \textbf{329}, 544-547 (2010).


\bibitem{SBL2013} S. Barraza-Lopez, A. A. Pacheco Sanjuan, Z. Wang, M. Vanevic, Solid State Commun. \textbf{166}, 70-75 (2013).

\bibitem{Qin2014} R. Qin, W. Zhu, Y. Zhang, and X. Deng, Nanoscale Res Lett. \textbf{9}, 521 (2014).




\bibitem{Wan2013} W. Wan, Y. Ge, F. Yang, Y. Yao, EPL. \textbf{104}, 36001 (2013).



\bibitem{Mohiuddin2009} T. M. G. Mohiuddin, A. Lombardo, R. R. Nair, A. Bonetti, G. Savini, R. Jalil, N. Bonini, D. M. Basko, C. Galiotis, N. Marzari, K. S. Novoselov, A. K. Geim, and A. C. Ferrari, Phys. Rev. B \textbf{79}, 205433 (2009).

\bibitem{Metzger2010} C. Metzger, S. Remi, M. Liu, S. V. Kusminskiy, A. H. Castro Neto, A. K. Swan and B. B. Goldberg, Nano Lett. \textbf{10}, 6–10 (2010).

\bibitem{Ding2010} F. Ding, H. Ji, Y. Chen, A. Herklotz, K. D\"{o}rr, Y. Mei, A. Rastelli and O. G. Schmidt, Nano Lett. \textbf{10}, 3453–3458 (2010).

\bibitem{Zabel2012} J. Zabel, R. R. Nair, A. Ott, T. Georgiou, A. K. Geim, K. S. Novoselov, and C. Casiraghi, Nano Lett. \textbf{12}, 617–621 (2012).

\bibitem{Ferralis2010} N. Ferralis, J. Mater. Sci. \textbf{45}, 5135 (2010).


\bibitem{Zhang2009} Yuanbo Zhang, Tsung-Ta Tang, Caglar Girit, Zhao Hao, Michael C. Martin, Alex Zettl,
Michael F. Crommie1, Y. Ron Shen and Feng Wang, Nature \textbf{459}, 820 (2009).



\bibitem{Huang2012} Mingyuan Huang, Hugen Yan, T. F. Heinz, and James Hone, Nano Lett. \textbf{10}, 4074-4079 (2010).

\bibitem{Li2012} Xiao Li, Rujing Zhang, Wenjian Yu, Kunlin Wang, Jinquan Wei, Dehai Wu, Anyuan Cao,
Zhihong Li, Yao Cheng, Quanshui Zheng, Rodney S. Ruoff and Hongwei Zhu, Scientific Reports \textbf{2}, 870 (2012).





\bibitem{me2013} J. A. Yan, R. Stein, D. M. Schaefer, X. Q. Wang, M. Y. Chou, Phys. Rev. B \textbf{88}, 121403(R) (2013).

\bibitem{Baroni2001} S. Baroni, S. de Gironcoli, and A. Dal Corso, Rev. Mod. Phys. \textbf{73}, 515-562 (2001).

\bibitem{pwscf} P. Giannozzi, \emph{et al.}, J. Phys. Condens. Matter \textbf{21}, 395502 (2009).
%


\bibitem{Troullier1991}  N. Troullier and J. L. Martins, Phys. Rev. B \textbf{43}, 1993-2006 (1991).


\bibitem{Bechstedt2012} F. Bechstedt, L. Matthes, P. Gori, and O. Pulci, App. Phys. Lett. \textbf{100}, 261906-261908 (2012).

\bibitem{Gonze1988} X. Gonze, B. Amadon, P.-M. Anglade, et al. Comput. Phys. Commun. \textbf{180}, 2582-2615 (2009).


\bibitem{Hybertsen1986} M. S. Hybertsen, and S. G. Louie, Phys. Rev. B \textbf{34}, 5390-5413 (1986).


\bibitem{Godby1988} R. W. Godby, M. Schl\"{u}ter, L. J. Sham, Phys. Rev. B \textbf{37}, 10159-10175 (1988).



\bibitem{Cheng2011} Y. C. Cheng, Z. Y. Zhu, G. S. Huang, and U. Schwingenschl\"{o}gl, Phys. Rev. B \textbf{83}, 115449 (2011).

\bibitem{note} We have also performed first-principles calculations including the spin-orbit coupling. We found that
the basic conclusions here are still valid, i.e., the compressive strain enhances the band gap opening in the band insulator regime, while the
tensile strain mitigates this band gap.


\end{thebibliography}
\end{document}